\documentclass[journal,twoside,web]{ieeecolor}
\usepackage{generic}
\usepackage{cite}
\usepackage{amsmath,amssymb,amsfonts}
\usepackage{algorithmic}
\usepackage{graphicx}
\usepackage{textcomp}
\usepackage{url}

\usepackage{amsthm}
\usepackage{cancel}
\def\BibTeX{{\rm B\kern-.05em{\sc i\kern-.025em b}\kern-.08em
    T\kern-.1667em\lower.7ex\hbox{E}\kern-.125emX}}
\theoremstyle{plain}
\newtheorem{theorem}{Theorem}

\newtheorem{corollary}{Corollary}
\newtheorem{proposition}{Proposition}

\theoremstyle{assumption}
\newtheorem{definition}{Definition}

\newtheorem{remark}{Remark}
\theoremstyle{goal}

\theoremstyle{assumption}

\begin{document}
\title{Exponentially Stable Adaptive Control Under Semi-PE Condition}
\author{Anton Glushchenko, \IEEEmembership{Member, IEEE}, and Konstantin Lastochkin
\thanks{Research was partly financially supported by Grants Council of the President of the Russian Federation (project MD-1787.2022.4).}
\thanks{A. I. Glushchenko is with V.A. Trapeznikov Institute of Control Sciences RAS, Moscow, Russia (phone: +79102266946; e-mail: aiglush@ipu.ru).}
\thanks{K. A. Lastochkin is with V.A. Trapeznikov Institute of Control Sciences RAS, Moscow, Russia (e-mail: lastconst@ipu.ru).}}

\maketitle

\begin{abstract}
A novel method of exponentially stable adaptive control to compensate for matched parametric uncertainty under a mild condition of semi-persistent excitation (s-PE) of a regressor with piecewise-constant rank and nullspace is proposed. It is based on the generalized dynamic regressor extension and mixing procedure developed earlier by the authors, does not require high adaptive gain or data stacks and ensures: 1) exponential convergence of the tracking error to zero and the parameter one to a bounded set when the regressor is s-PE, \linebreak 2) adjustable parameters transients of first-order type (each scalar parameter is adjusted using a separate first-order scalar differential equation), 3) alertness to change of the uncertainty parameters values, and 4) boundedness of all signals when the regressor is not s-PE. {The main salient feature of the proposed approach is that the exponential stability is guaranteed when the controller parameters estimates converge to the values that are indistinguishable from the true ones.} The results of numerical experiments fully support the theoretical analysis and demonstrate the advantages of the proposed method.
\end{abstract}


\section{Introduction}
\label{sec:introduction}

{\IEEEPARstart{P}{\textbf{osition}} \textbf{of Adaptation in the Control Theory.} Control systems design for technical plants requires analysis, formalization and compensation of disturbances of different nature. Both internal parametric and exogenous functional uncertainties have a crucial effect on the stability of the control system in {practical} applications. The internal ones are caused by a lack of {\it a priori} information about the system parameters, while the exogenous ones are introduced by external disturbances. If the majorant of the uncertainty is known (e.g., the admissible range of the system parameters variation, the maximum value of the external disturbance, etc.), the control system synthesis problem can be effectively solved using robust approaches: LMI synthesis \cite{b1}, $\mathcal{H}_{2}$ and $\mathcal{H}_{\infty}$ optimization \cite{b2}, sliding mode control \cite{b3}, methods of nonlinear robust control \cite{b4}. If the information about the system uncertainty is not available, or if the perturbations or uncertainties are large, then the control system should be designed with the help of adaptive control techniques \cite{b5}.

\textbf{Preview and Background.} The adaptive control systems are based on the duality principle, which implies simultaneous {\it exploration} of the system and {\it explotation} (adjustment) of the control law parameters to the current conditions \cite{b6}. To implement this principle, the parametric and functional uncertainties of the system are parameterized as ({\it i}) a regression equation with unknown parameters and a measurable regressor formed from the basis functions of the uncertainty {or ({\it ii}) using universal approximators}. The control law is chosen according to the {\it certainty equivalence} principle \cite{b7} again in the form of a regression with the same regressor, but with respect to the estimates of the unknown parameters. To overcome structural obstacles, if the control signal and uncertainty are in different equations, an iterative backstepping procedure is applied \cite{b8}.

The key step of all adaptive systems design is the derivation of the adaptive law to obtain the estimates of the unknown parameters. There are \cite{b5} two classical ways to solve this problem: 1) {\it{Lyapunov}} based design; 2) {\it{Modular}} (Estimation) based design. The first way is to derive the adaptive law from the first derivative of a properly chosen Lyapunov function. The second way implies that the parameterized uncertainty in static form is obtained from the plant equations and then the gradient descent or least squares method is applied to derive the adaptation algorithm. Both procedures require the unknown parameters to be time-invariant. In case when all disturbances are taken into consideration and included into the parameterization of the uncertainty, the resulting adaptive laws guarantee both the boundedness of all closed-loop system signals and the asymptotic stability of the equilibrium of the tracking error (difference between the output of the plant and the reference model). Estimates of the unknown parameters do not necessarily coincide with their true values, but usually belong to stabilizing manifolds \cite{b9, b10}. If, additionally, the strict condition of {\it{persistent excitation}} (PE) is satisfied, then the exponential convergence of estimates to the ground-truth values and exponential stability of the tracking error are guaranteed \cite{b11}. Exponentially fast convergence of plant state to the reference one is a very important property because it provides dynamic recovery of desired transient response and stability properties under the condition of exogenous disturbance \cite{b12}.

\textbf{Motivation.} When the exponential stability condition is not met, and the measurement noise or disturbances that are not included into the parameterization affect the plant, the system suffers from the unbounded drift of the adjustable parameters \cite{b13}, and uniform improvement of the tracking error transients quality is impossible without high gain adaptation \cite{b14, b15}.

Considerable attention of the control science community has been focused on these problems in recent decades. The first proposed solutions were robust modifications of basic adaptive laws (dead zone \cite{b16}, projection operator \cite{b17}, leakage methods \cite{b13, b18}, low frequency learning \cite{b19}, optimal control modification \cite{b20} etc.). They ensure boundedness of the adjustable parameters under external disturbances, but do not guarantee the parameter error exponential convergence without the regressor persistent excitation.

Then the excitation injection approaches \cite{b20i1,b20i2,b20i3} have been proposed, which allows one to design a plant input signal in such a way that the PE condition is ensured to be met. This signal is usually a sum of harmonic functions. In many practical applications such injection can lead to damage of actuators, energy consumption increase, and cause situations when the plant functions out of standard operating mode.

To overcome above-mentioned obstacles, more recent solutions are mostly aimed at the exponential stability condition relaxation. Since the concurrent learning proposed in the disseminating paper \cite{b21}, many different approaches \cite{b22,b23,b24,b25,b26,b27,b28,b29,b30,b31}, \cite{b38,b39,b40,b41} to guarantee parameter convergence and exponential stability under relaxed regressor excitation requirements have been proposed in journals and conferences proceedings. A comparative simulation and detailed discussion on some of them are presented in \cite{b31, b32}.

In \cite{b22,b23,b24,b25,b26,b27,b28,b29,b30,b31}, inertial dynamic filters of static regressions or algorithms to form data stacks are used. Such ways to store the information about the uncertainty and regressor values in the "memory" of the filters or special stacks allows one to relax PE requirement to the ones of the {\it{finite}} or {\it{initial excitation}}  (FE or IE). The PE condition is equivalent to the one of identifiability of uncertainty parameters over the entire time axis, whereas the weaker FE or IE conditions – only over a finite time range \cite{b30, b33}. However, the composite laws \cite{b22,b23,b24,b25,b26,b27,b28,b29,b30,b31} provide unsatisfactory tracking performance in the case of even small changes of the uncertainty parameters. The aim to compensate for the uncertainty with time-varying parameters is often the main motivation of the adaptive systems application. Nowadays, however, apart from some {\it ad hoc} quick fixes \cite{b31}, \cite{b34,b35,b36,b37}, no effective long term solutions have been proposed to make the composite laws capable of compensation for the uncertainty with time-varying or piecewise-constant parameters.

In \cite{b38} an adaptive law is premultiplied with an exponential function to obtain time-varying adaptive gain, which compensates for the regressor excitation vanishing and provides the exponential convergence of the tracking error with independent of the initial conditions and user-assignable rate. In \cite{b39, b40} a least-square-based adaptive law is proposed, which provides boundedness of a time-varying adaptive gain and exponential convergence of the tracking and parameter errors to a compact set under the mild excitation requirements. As thoroughly discussed in \cite{b15}, the high-gain adaptation in \cite{b38,b39,b40} may result in noise amplification, unmodelled dynamics instability and other bursting phenomena, which is dramatically critical for a practical scenario \cite{b13}.

Prescribed Performance Control (PPC) methods \cite{b41, b42, b43} ensure semi-global exponential stability with a desired rate of convergence of the tracking error to a prescribed set. However, ({\it{i}}) the initial system conditions should be appropriately known to set the PPC parameters, ({\it{ii}}) the used diffeomorphisms are defined via exponential functions, and the adaptive laws use high gain; ({\it{iii}}) convergence of the tracking error to zero may not be achieved in order to avoid the potential singularity issue.

In \cite{b44, b45}, another approach to guarantee the exponential stability of adaptive control systems is proposed. Instead of PE or FE/IE, the exponential stability condition is the {\it {semi-persistent excitation}} (s-PE or the lack of PE) of the regressor, which can be equivalent to the condition of partial identifiability of the uncertainty parameters \cite{b46}. This condition is satisfied with rank one as long as at least one element of the regressor is globally bounded away from zero. Therefore, \linebreak s-PE is strictly weaker than PE/FE/IE conditions and satisfied with rank one even when all regressor elements are linearly dependent \cite{b44}.

First of all, in \cite{b44, b45} it is shown that, when the regressor is s-PE, the exponential convergence of the tracking error between the plant and reference model states can be ensured by the exponential convergence of the identification error of some new {\it indistinguishable parameters}. In the general case, the new parameters do not coincide with the uncertainty original parameters and are their projections on the nullspace of the regressor. The authors of \cite{b45} have derived modified gradient and recursive least squares adaptive laws that guarantee exponential convergence of the identification error of the indistinguishable parameters to zero, which additionally ensures the boundedness of the identification error of the uncertainty original parameters.

The main advantage of the approach \cite{b44, b45}, compared to the well-known solutions \cite{b22,b23,b24,b25,b26,b27,b28,b29,b30,b31}, is that it uses neither inertial filters, nor off-line procedures to obtain, monitor and update the data stack. Because of this, not only do the solutions from \cite{b44, b45} provide exponential stability of the tracking error and robustness of the closed-loop adaptive system to disturbances, but also they have tracking capability of the uncertainty time-varying parameters under sufficient excitation. However, methods from \cite{b44, b45}, firstly, require {\it a priori} information about the rank of the regressor s-PE and assume that such rank is time-invariant, and secondly, the proposed adaptive laws ensure monotonicity of the parameter error norm only, while the transient of each certain unknown parameter estimate may suffer from significant oscillations.

A recently developed identification law \cite{b47} based on the generalized dynamic regressor extension and mixing (G-DREM) procedure uses a similar concept \cite{b44, b45}.

The conventional DREM \cite{b48} consists of two key steps: regressor {\it{extension}} and {\it{mixing}}. In the extension step, using special stable minimum-phase operators, the original regressor is converted into a quadratic one. In the mixing step, the quadratic regressor is transformed into a scalar one by the multiplication of the extended regression with the adjoint matrix of the extended regressor. The regression obtained after mixing is a set of scalar equations with respect to the original unknown parameters. As for the basic DREM procedure, the scalar regressor is bounded away from zero and preserves excitation of the original regressor only when the FE condition is met \cite{b49, b50}. In the G-DREM procedure \cite{b47} a new step of {\it dynamic regularization} based on the eigenvalue decomposition is introduced. Such step allows one to obtain: 1) a regression with respect to indistinguishable parameters that are projections of the original parameters on the nullspace of the regressor; 2) a scalar regressor that is separated from zero when the weaker semi-finite excitation (s-FE) condition is met.

The identification law derived on the basis of the G-DREM procedure \cite{b47} has all advantages of modified gradient descent and recursive least squares laws \cite{b44, b45} and in addition:
\begin{enumerate}
    \item[--] assumes that the rank of s-PE of the regressor is piecewise-constant;
    \item[--] if the regressor is s-PE with a constant rank, it guarantees monotonicity of the estimation errors transients of the indistinguishable parameters vector elements;
    \item[--] ensures the adjustable parameters transients of first-order type (each scalar parameter is adjusted using a separate first-order scalar differential equation).
\end{enumerate}

\textbf{Contribution and organization of the paper.} In \cite{b47}, the properties of the adaptive law based on the G-DREM procedure have been proved for the problem of identification of the static linear regression equation unknown parameters. This study is to take one step forward with respect to the results of \cite{b47} and apply the new adjustment algorithm to the problem of adaptive control. The main purpose of this study is to develop an adaptive control system based on the G-DREM procedure \cite{b47} for a class of systems with matched uncertainty.

In the first step of the adaptive system synthesis, following the modular based design \cite{b5}, a control law is chosen for the system with uncertainty, which ensures the boundedness of all signals for any bounded estimates of the uncertainty parameters. In the second step, a static regression model with respect to the unknown uncertainty parameters is obtained on the basis of the measured signals. In the third step, a G-DREM procedure \cite{b47} is applied to the obtained regression, and a regression equation with a scalar regressor with respect to the new indistinguishable parameters is derived. In the fourth step, using the obtained regression, the adaptive law is introduced and the stability of the control system is analyzed.

Generally, the novelty and contributions of the proposed method that distinguishes it from most previous results are summarized as follows:
\begin{enumerate}
    \item[\textbf{C1}] a new method of exponentially stable adaptive control to compensate for matched parametric uncertainty under s-PE condition is proposed. In contrast to \cite{b44, b45}, the rank of the s-PE condition is assumed to be piecewise-constant. Contrary to the composite laws \cite{b22,b23,b24,b25,b26,b27,b28,b29,b30,b31}, not only does the adaptive control system guarantee exponential stability under the relaxed regressor excitation requirement, but also it provides {\it alertness} to the uncertainty parameters variation. Unlike solutions from \cite{b38,b39,b40,b41,b42, b43}, an adaptive law does not require high adaptive gain to provide closed-loop exponential stability;
    \item[\textbf{C2}] the boundedness of all signals is guaranteed regardless of the exponential stability condition;
    \item[\textbf{C3}] the quality of the adjustable parameters transients is of first-order type (each scalar parameter is adjusted using a separate first-order scalar differential equation).

\end{enumerate}

The remainder of the paper is organized as follows. Notation employed in this paper is given below. Section II briefly discusses the properties and advantages of the identification law \cite{b47}. The problem of the exponentially stable adaptive control for a class of systems with matched parametric uncertainty is stated in Section III. The main result is elucidated in Section IV. Section V presents the results of numerical experiments. The paper is wrapped-up with conclusion in Section VI.

\textbf{Notation and Definition.} Further the following notation is used: $\left| . \right|$ is the absolute value or, whenever it follows from the context, the power of the set, $\left\| . \right\|$ is the suitable norm of $(.)$, ${\lambda _{min }}\left( . \right)$ and ${\lambda _{max }}\left( . \right)$  are the matrix minimum and maximum eigenvalues respectively, ${I_{n \times n}}$ is an identity $n\!\!\! \times \!\!\! n$ matrix, ${0_{n \times n}}$ is a zero $n \!\! \times \!\! n$ matrix, $det\{.\}$ stands for a matrix determinant, $adj\{.\}$ represents an adjoint matrix, ${\mathbb{L}_\infty }$ is the space of all essentially bounded functions, {$\perp$ denotes orthogonality}, $\mathfrak{H}\left[ . \right]{\rm{:}} = {\textstyle{1 \over {p + {k_0}}}}\left[ . \right]$ is the linear stable minimum-phase operator with ${k_0} > 0$, ${p}{\rm{:}} = {\textstyle{d \over {dt}}}$ is a derivative operator.

\begin{definition}\label{definition1}
The regressor $\overline \varphi \left( t \right) \in {\mathbb{R}^n}$ is persistently exciting $\left( {\overline \varphi \left( t \right) \in {\rm{PE}}} \right)$  if $\forall t \ge {t_0} \ge 0$, $ \exists T > 0$ and $\alpha  > 0$ such that the following inequality holds:
\begin{equation}\label{eq1}
{\lambda _{{{min}}}}\left\{ {\int\limits_t^{t + T} {\overline \varphi \left( \tau  \right){{\overline \varphi }^{\rm{T}}}\left( \tau  \right)d} \tau } \right\} \ge \alpha ,
\end{equation}
where $\alpha  > 0$ is the excitation level.
\end{definition}
\begin{definition}\label{definition2}
The regressor $\overline \varphi \left( t \right) \in {\mathbb{R}^n}$ is semi-persistently exciting $\left( {\overline \varphi \left( t \right) \in {\rm{s\text{-}PE}}} \right)$ with full rank $0 < r < n$ if $\forall t \ge {t_0} \ge \linebreak \ge 0$, $\exists T > 0$ and $0 < \underline{\alpha} \le \overline \alpha$ such that $\forall i \in \left\{ {1, \ldots {\rm{,}}r} \right\}$  the following inequality holds:
\begin{equation}\label{eq2}
\underline{\alpha} \le {\lambda _i}\left\{ {\int\limits_t^{t + T} {\overline \varphi \left( \tau  \right){{\overline \varphi }^{\rm{T}}}\left( \tau  \right)d} \tau } \right\} \le \overline \alpha,
\end{equation}
where $0 < \underline{\alpha} \le \overline \alpha$ is the excitation level.
\end{definition}

The properties of $\varphi \left( t \right) = \mathfrak{H}\left\{{\overline \varphi \left( t \right){{\overline \varphi }^{\rm{T}}}\left( t \right)} \right\} \in {\mathbb{R}^{n \times n}}$ are copied from \cite{b47, b50, b51} for the cases when the inequality \eqref{eq1} or \eqref{eq2} holds for $\overline \varphi \left( t \right) \in {\mathbb{R}^n}$.
\begin{corollary}
    $\overline \varphi \left( t \right) \in {\rm{PE}} \Leftrightarrow \forall t \ge kT{\rm{\;}}{\lambda _{{\rm{min}}}}\left( t \right) > \mu {\rm{.}}$
\end{corollary}
\begin{corollary}
    $\overline \varphi \left( t \right) \in {\rm{s\text{-}PE}} \Leftrightarrow \forall t \ge kT{\rm{\;}}\forall i \in \overline {1,{\rm{\;}}r} {\rm{\;}}{\lambda _i}\left( t \right) > \mu .$
\end{corollary}

Here $k \ge 1$ is a positive integer number, $\mu  > 0$ is an eigenvalue lower bound, ${\lambda _i}\left( t \right)$ is the $i^{th}$ eigenvalue of the regressor $\varphi \left( t \right)$, ${\lambda _{{{min}}}}\left( t \right) = \mathop {{\rm{min}}}\limits_{1 \le i \le n - \overline r} {\lambda _i}\left( t \right)$ denotes minimal non-zero eigenvalue of the regressor $\varphi \left( t \right)$, $\overline r = n - r$ stands for the rank deficiency

The proof of Corollary 1 is given in \cite{b50, b51}, whereas the proof of Corollary 2 could be derived in the same way.

Based on the definition of the eigenvalue decomposition of the positive semidefinite time-invariant matrix from \cite{b2}, the following definition of the eigenvalue decomposition of $\varphi \left( t \right) \in {\mathbb{R}^{n \times n}}$ is introduced.
\begin{definition}
    The eigenvalue decomposition of the regressor $\varphi \left( t \right) \in {\mathbb{R}^{n \times n}}$ with the piecewise-constant rank $r\left( t \right) \le n$ is defined as the following multiplication:
    \begin{equation}\label{eq3}
\begin{array}{c}
{V^{\rm{T}}}\left( t \right)\varphi \left( t \right)V\left( t \right) = \left[ \begin{array}{l}
V_1^{\rm{T}}\left( t \right)\\
V_2^{\rm{T}}\left( t \right)
\end{array} \right]\varphi \left( t \right)\left[ {{V_1}\left( t \right){\rm{ }}{V_2}\left( t \right)} \right] = \\
 = \Lambda \left( t \right) = \left[ {\begin{array}{*{20}{c}}
{{\Lambda _1}\left( t \right)}&{{0_{r\left( t \right) \times \overline r\left( t \right)}}}\\
{{0_{\overline r\left( t \right) \times r\left( t \right)}}}&{{0_{\overline r\left( t \right)\times {\overline r\left( t \right)}}}}
\end{array}} \right]{\rm{,\;}}{\Lambda _1}\left( t \right) > 0,
\end{array}
\end{equation}
where  ${V_1}\left( t \right) \in {\mathbb{R}^{n \times r\left( t \right)}}$ stands for a time-varying orthonormal basis of the eigenspace of $\varphi \left( t \right)$,  ${V_2}\left( t \right) \in {\mathbb{R}^{n \times \overline r\left( t \right)}}$ is a time-varying orthonormal basis of the nullspace of $\varphi \left( t \right)$, ${\lambda _1}\left( t \right) \ge \linebreak \ge\! {\lambda _2}\left( t \right)\! \ge \!  \cdots \! \ge \!{\lambda _{r\left( t \right)}}\left( t \right) > 0$  are nonzero eigenvalues of $\varphi \left( t \right)$. 
\end{definition}

\section{Preliminaries}

Before proceeding directly to the adaptive control problem statement, the results obtained in \cite{b47} are briefly summarized.

The identification problem of time-invariant parameters of the linear regression equation is considered:
    \begin{equation}\label{eq4}
\forall t \ge t_0^ + {\rm{\;}}z\left( t \right) = {\overline \varphi ^{\rm{T}}}\left( t \right)\theta {\rm{,}}
\end{equation}
where $\overline \varphi \left( t \right) \in {\mathbb{L}_\infty } \cap {\mathbb{R}^n}$, $z\left( t \right) \in {\mathbb{L}_\infty } \cap \mathbb{R}$ are the measurable bounded regressor and function (regressand), $\theta  \in {\mathbb{R}^n}$ stands for the unknown time-invariant $\left( {\dot \theta  \equiv 0} \right)$ and bounded $\left( {\left\| \theta  \right\| \le {\theta _{{{max}}}}} \right)$ parameters\footnote{All results are also valid for: $ z\left( t \right) \in {\mathbb{R}^{q \times p}}{\rm{,\;}}\overline \varphi \left( t \right) \in {\mathbb{R}^{n \times q}}{\rm{,\;}}\theta  \in {\mathbb{R}^{n \times p}}.$}.		

Using the dynamic extension
    \begin{equation}\label{eq5}
\varphi \left( t \right){\rm{:}} = \mathfrak{H}\left\{ {\overline \varphi \left( t \right){{\overline \varphi }^{\rm{T}}}\left( t \right)} \right\}{\rm{,\;}}y\left( t \right){\rm{:}} = \mathfrak{H}\left\{ {\overline \varphi \left( t \right)z\left( t \right)} \right\}{\rm{,}}
\end{equation}
dynamic regularization
    \begin{equation}\label{eq6}
\begin{array}{c}
y\left( t \right) = \varphi \left( t \right)\theta  = \\
 = V\left( t \right)\Lambda \left( t \right){V^{\rm{T}}}\left( t \right)\theta  \pm V\left( t \right)\Xi \left( t \right){V^{\rm{T}}}\left( t \right)\theta  = \\
 = V\left( t \right)\overline \Lambda \left( t \right){V^{\rm{T}}}\left( t \right)\theta  - V\left( t \right)\Xi \left( t \right){V^{\rm{T}}}\left( t \right)\theta  = \\
 = \Phi \left( t \right)\theta  - V\left( t \right)\Xi \left( t \right){V^{\rm{T}}}\left( t \right)\theta  = \Phi \left( t \right)\Theta \left( t \right),\\
\Xi \left( t \right) = \overline \Lambda \left( t \right) - \Lambda \left( t \right){\rm{,\;}}\Theta \left( t \right) = \theta  - {V_2}\left( t \right)V_2^{\rm{T}}\left( t \right)\theta {\rm{,}}\\
\overline \Lambda \left( t \right){\rm{:}} = \left\{ \begin{array}{l}
0,{\rm{\;if\;diag}}\left\{ {{{\overline \lambda }_1}\left( t \right){\rm{,}} \ldots {\rm{,}}{{\overline \lambda }_n}\left( t \right)} \right\} = \varepsilon {I_{n \times n}}\\
{\rm{diag}}\left\{ {{{\overline \lambda }_1}\left( t \right){\rm{,}} \ldots {\rm{,}}{{\overline \lambda }_n}\left( t \right)} \right\}{\rm{,\;otherwise}}
\end{array} \right.\\
{{\overline \lambda }_i}\left( t \right){\rm{:}} = \left\{ \begin{array}{l}
{\lambda _i}\left( t \right){\rm{,\;if\;}}{\lambda _i}\left( t \right) \ge \overline \varepsilon \\
\varepsilon {\rm{,\;if\;}}{\lambda _i}\left( t \right) < \overline \varepsilon 
\end{array} \right.{\rm{,\;}}i = \overline {1,n} {\rm{,  }}
\end{array}
\end{equation}
and the mixing procedures:
    \begin{equation}\label{eq7}
\Upsilon \left( t \right){\rm{:}} = adj\left\{ {\Phi \left( t \right)} \right\}y\left( t \right){\rm{,\;}}\omega \left( t \right){\rm{:}} = det \left\{ {\Phi \left( t \right)} \right\},
\end{equation}
a measurable regression equation with scalar regressor is obtained:
    \begin{equation}\label{eq8}
\Upsilon \left( t \right) = \omega \left( t \right)\Theta \left( t \right){\rm{,}}
\end{equation}
where $\omega \left( t \right) \ge {\rm{min}}\left\{ {\lambda _{{\rm{min}}}^n\left( t \right){\rm{,\;}}{\varepsilon ^n}} \right\} \ge {\rm{min}}\left\{ {\mu^n{\rm{,\;}}{\varepsilon ^n}} \right\}  > 0$, the new unknown parameters $\Theta\left( t \right)$ meet the indistinguishability condition
\begin{center}
    $
    {\overline \varphi ^{\rm{T}}}\left( t \right)\theta -{\overline \varphi ^{\rm{T}}}\left( t \right)\Theta\left( t \right)=0,
    $
\end{center}
$\varepsilon > 0$ is a parameter that defines the value of virtual eigenvalues, $\overline \varepsilon  \ge 0$ denotes a parameter that defines the amplitude of the eigenvalues of $\varphi(t)$, which are considered to be equivalently equaled to zero in the presence of computational errors and external disturbances.

Based on \eqref{eq8}, the estimation law is introduced:
\begin{equation}\label{eq9}
\begin{array}{c}
\dot {\hat {\theta}} \left( t \right) =  - \gamma \left( t \right)\omega \left( t \right)\left( {\omega \left( t \right)\hat \theta \left( t \right) - \Upsilon \left( t \right)} \right) = \\
 =  - \gamma \left( t \right){\omega ^2}\left( t \right)\underbrace {\left( {\hat \theta \left( t \right) - \Theta \left( t \right)} \right)}_{\tilde \Theta \left( t \right)}{\rm{,\;}}\hat \theta \left( {t_0^ + } \right) = {\theta _0}{\rm{, }}\\
\gamma \left( t \right){\rm{:}} = \left\{ \begin{array}{l}
{\gamma _1}{\rm{,\;if\;}}\omega \left( t \right) \le {\rm{min}}\left\{ {\mu^n{\rm{,\;}}{\varepsilon ^n}} \right\}\\
{\textstyle{{{\gamma _0}} \over {{\omega ^2}\left( t \right)}}}{\rm{,\;otherwise}}
\end{array} \right.
\end{array}
\end{equation}
where ${\gamma _0} > 0,{\rm{\;}}{\gamma _1} > 0$ are arbitrary parameters of the estimation law, $\tilde \Theta \left( t \right) \in {\mathbb{R}^n}$ denotes the identification error.

As it has been proved in \cite{b47}, with respect to the errors $\tilde \Theta \left( t \right){\rm{,\;}}\tilde z\left( t \right) = \hat z\left( t \right) - z\left( t \right) = {\overline \varphi ^{\rm{T}}}\left( t \right)\hat \theta \left( t \right) - z\left( t \right){\rm{,\;}}\tilde \theta \left( t \right) = \linebreak = \hat \theta \left( t \right) - \theta \left( t \right)$, the law \eqref{eq9} ensures the following properties:
\begin{enumerate}
    \item [\textbf{I.}] \begin{enumerate}
        \item[1)] $\overline \varphi \left( t \right) \in {\rm{s\text{-}PE}} \Rightarrow \mathop {{\rm{lim}}}\limits_{t \to \infty } \left\| {\tilde \theta \left( t \right)} \right\| \le {\theta _{{{max}}}}\left( {{\rm{exp}}} \right).$
        \item[2)] $\overline \varphi \left( t \right) \in {\rm{PE}} \Rightarrow \left\{ \begin{array}{l}
\mathop {{\rm{lim}}}\limits_{t \to \infty } \left\| {\tilde \theta \left( t \right)} \right\| = 0{\rm{ }}\left( {{\rm{exp}}} \right){\rm{;}}\\
\mathop {{\rm{lim}}}\limits_{t \to \infty } \left| {\tilde z\left( t \right)} \right| = 0{\rm{ }}\left( {{\rm{exp}}} \right).
\end{array} \right.$
    \end{enumerate}
    \item [\textbf{II.}]  If there exists the decomposition \eqref{eq3} with the time-invariant matrix  ${V_2}\left( t \right) \equiv {V_2}$ for $\varphi \left( t \right) \in {\mathbb{R}^{n \times n}}$ with the time-invariant rank $r\left( t \right) \equiv r < n{\rm{,\;}}\overline r\left( t \right) \equiv \overline r > 0$, then, in addition to the properties \textbf{I}, the law \eqref{eq9} also provides:
    \begin{enumerate}
        \item [1)] $\left| {{{\tilde \Theta }_i}\left( {{t_a}} \right)} \right| \le \left| {{{\tilde \Theta }_i}\left( {{t_b}} \right)} \right|{\rm{ }}\forall {t_a} \ge {t_b}{\rm{,\;}}\forall i \in \left\{ {1,{\rm{ }}n} \right\}.$
    \item[2)] $\overline \varphi \left( t \right) \in {\rm{s\text{-}PE}} \Rightarrow \left\{ \begin{array}{l}
\mathop {{\rm{lim}}}\limits_{t \to \infty } \left\| {\tilde \Theta \left( t \right)} \right\| = 0{\rm{ }}\left( {{\rm{exp}}} \right){\rm{;}}\\
\mathop {{\rm{lim}}}\limits_{t \to \infty } \left| {\tilde z\left( t \right)} \right| = 0{\rm{ }}\left( {{\rm{exp}}} \right).
\end{array} \right.$
    \end{enumerate}
    \item [\textbf{III.}] If $r\left( t \right)$ is a piecewise-constant function:
 \begin{equation}\label{eq10}
\forall t \ge t_0^ + {\rm{\;}}r\left( t \right) = \sum\limits_{{j_r} = 1}^{j_r^{max }} {{\Delta _{{j_r}}}h\left( {t - {t_{{j_r}}}} \right)} {\rm{,}}
\end{equation}   
there exists the decomposition \eqref{eq3} with the piecewise-constant matrix ${V_2}\left( t \right)$:
 \begin{equation}\label{eq11}
\forall t \ge t_0^ + {\rm{\;}}{V_2}\left( t \right) = \sum\limits_{{j_V} = 1}^{j_V^{max }} {{\Delta _{{j_V}}}h\left( {t - {t_{{j_V}}}} \right)} {\rm{,}}
\end{equation}  
the unknown parameters $\Theta \left( t \right)$ are the piecewise-constant functions:
 \begin{gather*}
     \Theta \left( t \right) = \theta  + \sum\limits_{q = 1}^j {{\Delta _q}h\left( {t - {t_q}} \right)} .
 \end{gather*}
and if $j \le {j_{{\rm{max}}}} < \infty {\rm{}}$,  $\overline \varphi \left( t \right) \in {\rm{s\text{-}PE}}$ with rank $r\left( t \right) \ge 1$, then in addition to the properties from \textbf{I}, it holds that:
 \begin{gather*}
\left\{ \begin{array}{l}
\mathop {{\rm{lim}}}\limits_{t \to \infty } \left| {\tilde z\left( t \right)} \right| = 0{\rm{ }}\left( {{\rm{exp}}} \right){\rm{ }}\\
\mathop {{\rm{lim}}}\limits_{t \to \infty } \left\| {\tilde \Theta \left( t \right)} \right\| = 0{\rm{ }}\left( {{\rm{exp}}} \right)
\end{array} \right..
 \end{gather*}
\item[\textbf{IV.}] If the premises of \textbf{II} and the following conditions hold\footnote{Without loss of generality, it is assumed that the first $n-p$ columns of the regressor $\varphi \left( t \right) = \left[ {{\varphi _1}\left( t \right) \ldots {\varphi _i}\left( t \right) \ldots {\varphi _n}\left( t \right)} \right]$ are linearly dependent (in case $\overline r\left( t \right) > 0$ such form can always be obtained by columns permutation).}:
 \begin{gather*}
\begin{array}{c}
\overline \varphi \left( t \right) \in {\rm{s\text{-}PE}}{\rm{,\;}}n > 2,\\
\sum\limits_{i = 1}^{n - p} {{w_i}{\varphi _i}\left( t \right)}  + \sum\limits_{j = n - p + 1}^n {{w_j}{\varphi _j}\left( t \right)}  = {0_n}{\rm{,\;}}{w_i} \ne 0,{\rm{\;}}{w_j} = 0,
\end{array}
 \end{gather*}
 then $\exists M \subset \left\{ {1,...,n} \right\}{\rm{,\;}}\left| M \right| = p{\rm{,\;}}\forall i \in M{\rm{,\;}}{\Theta _i} = {\theta _i}.$
\end{enumerate}

Here we use the following notation: ${t_{{j_r}}}$ is the time instant of the regressor rank change, ${\Delta _{{j_r}}}$ stands for the amplitude of rank change at the time instant ${t_{{j_r}}}$, ${t_{{j_V}}}$ denotes the time instant of the basis ${V_2}\left( t \right)$ change, ${\Delta _{{j_V}}} \in {\mathbb{R}^{n \times \overline r\left( t \right)}}$ is the amplitude of ${V_2}\left( t \right)$ change, $h\left( {t - {t_{{j_r}}}} \right){\rm{,\;}}h\left( {t - {t_{{j_V}}}} \right)$ are the unit step (Heaviside) functions, ${t_j} \in \left\{ {{t_{{j_r}}}{\rm{,\;}}{t_{{j_V}}}\left| {{j_r} \in \mathbb{N}{\rm{,\;}}{j_V} \in \mathbb{N}} \right.} \right\}$ are the time instants of the parameters $\Theta \left( t \right)$ change, $\left\| {{\Delta _j}} \right\| \le {\Delta _{{\rm{max}}}}$ stands for the bounded value of the parameters change.

The G-DREM law \eqref{eq9}, in contrast to the well-known:\linebreak gradient descent
\begin{center}
    $
    \dot {\hat {\theta}} \left( t \right) =  - \Gamma \overline \varphi \left( t \right)\left( {z\left( t \right) - {{\overline \varphi }^{\rm{T}}}\left( t \right)\hat \theta \left( t \right)} \right){\rm{,\;}}\Gamma  = {\Gamma ^{\rm{T}}} > 0,
    $
\end{center}
recursive least squares
\begin{center}
    $
    \begin{array}{c}
\dot {\hat {\theta}} \left( t \right) =  - \Gamma \overline \varphi \left( t \right)\left( {z\left( t \right) - {{\overline \varphi }^{\rm{T}}}\left( t \right)\hat \theta \left( t \right)} \right){\rm{,}}\\
\dot \Gamma  = {\lambda _f}\Gamma  - \Gamma \overline \varphi \left( t \right){{\overline \varphi }^{\rm{T}}}\left( t \right)\Gamma {\rm{,\;}}{\lambda _f} > 0,
\end{array}
    $
\end{center}
and DREM 
\begin{center}
    $
    \dot {\hat {\theta}} \left( t \right) =  - \gamma adj\left\{ {\varphi \left( t \right)} \right\}\left( {y\left( t \right) - \varphi \left( t \right)\hat \theta \left( t \right)} \right){\rm{,\;}}\gamma  > 0
    $
\end{center}
estimation laws, guarantees the global {\it{exponential convergence}} of the tracking error $\tilde z\left( t \right)$ to zero when $\overline \varphi \left( t \right) \in {\rm{s\text{-}PE}}$, the parameter error $\tilde \theta \left( t \right)$ to the bounded set with the bound ${\theta _{{{max}}}}$, and the first-order type transients of the estimates $\hat \theta \left( t \right)$ (each scalar parameter ${\hat \theta _i}\left( t \right)$ is adjusted using a separate first-order scalar differential equation). Moreover, if the premises of part \textbf{IV} hold, then the law \eqref{eq9} ensures {\it{partial identifiability}} of elements of the original unknown parameters vector.

The main aim of this study is to apply the law \eqref{eq9} to the problem of model reference adaptive control.

\begin{remark}
The main difficulty of the practical implementation of law \eqref{eq9} is the need to choose the constant $\bar \varepsilon $, which is used in the condition of substitution of eigenvalues $\varphi \left( t \right)$ that are sufficiently close to zero with $\varepsilon$ .
\end{remark}

\section{Problem Statement}

A generalized class of continuous linear time-invariant systems with matched uncertainty is considered:
\begin{equation}\label{eq12}
\dot x\left( t \right) = Ax\left( t \right) + B\left( {u\left( t \right) - {\theta ^{\rm{T}}}\Psi \left( {x{\rm{,\;}}t} \right)} \right){\rm{,\;}}
x\left( {t_0^ + } \right) = {x_0}{\rm{,\;}}
\end{equation} 
where $x\left( t \right) \in {\mathbb{R}^n}$ is a state vector with the unknown initial conditions ${x_0}$, ${u\left( t \right) \in {\mathbb{R}^m}}$ is a control vector, $A \in {\mathbb{R}^{n \times n}}$ is a known state matrix, $B \in {\mathbb{R}^{n \times m}}$ is a known input matrix, $t_0^+$ is a known initial time instant. The pair $\left( {A,{\rm{\;}}B} \right)$ is controllable, $\forall t > t_0^ +$ the vector $\Psi \left( {x{\rm{,\;}}t} \right) \in {\mathbb{R}^p}$ is measurable, the matrix $\theta  \in {\mathbb{R }^{p \times m}}$  is unknown, but bounded $\left\| \theta  \right\| \le {\theta _{max }}$.\footnote{ Without any confusion further we use notation $\Psi \left( {x{\rm{,\;}}t} \right) \equiv \Psi \left( t \right)$.}

The required control quality is defined with the help of a reference model with ${x_{ref}}\left( {t_0^ + } \right) = {x_{0ref}}$:
\begin{equation}\label{eq13}
\begin{array}{c}
{{\dot x}_{ref}}\left( t \right) = {A_{ref}}{x_{ref}}\left( t \right) + {B_{ref}}{z_{cmd}}\left( t \right){\rm{,}}\\
\end{array}
\end{equation} 
where ${x_{ref}}\left( t \right) \in {{{\mathbb{R}}} ^n}$ is a reference model state vector, ${x_{0ref}}$ is a vector of initial conditions, ${z_{cmd}}\left( t \right) \in {{ {\mathbb{ R}} } ^m}$ is a reference signal, ${A_{ref}} = A + B{K_x} \in {{{\mathbb{R}}} ^{n \times n}}$ is a Hurwitz state matrix of the reference model, ${B_{ref}} = B{K_r} \in {{ {\mathbb{R}} } ^{n \times m}}$ is a reference model input matrix.

The control law is chosen as:
\begin{equation}\label{eq14}
\begin{array}{c}
u\left( t \right) = {u_{bl}}\left( t \right) + {u_{ad}}\left( t \right) + {u_{nd}}\left( t \right){\rm{,}}\\
{u_{bl}}\left( t \right) = {K_x}x\left( t \right) + {K_r}{z_{cmd}}\left( t \right){\rm{, }}\\{u_{ad}}\left( t \right) = {{\hat \theta }^{\rm{T}}}\left( t \right)\Psi \left( t \right){\rm{, }}\\
{u_{nd}}\left( t \right) =  - \kappa {B^{\rm{T}}}P{e_{ref}}\left( t \right){\Psi ^{\rm{T}}}\left( t \right)\Psi \left( t \right){\rm{,}}
\end{array}
\end{equation} 
where ${K_x} =  - {B^{\rm{\dag }}}\left( {A - {A_{ref}}} \right) \in {\mathbb{R}^{m \times n}}$ is a matrix of feedback parameters, ${K_r} = {B^{\rm{\dag }}}{B_{ref}} \in {\mathbb{R}^{m \times m}}$ is a matrix of feedforward parameters, $\hat \theta \left( t \right) \in {\mathbb{R}^{p \times m}}$ is a matrix of adjustable parameters, $\kappa  > 0$ is a parameter of the nonlinear damping term, ${e_{ref}}\left( t \right) = x\left( t \right) - {x_{ref}}\left( t \right)$ is a tracking error, $P$ is matrix obtained as a solution of the Lyapunov equation 
\begin{gather*}
    A_{ref}^{\rm{T}}P + P{A_{ref}} =  - Q{\rm{,\;}}Q=Q^{\rm T} > 0.
\end{gather*}

The baseline term ${u_{bl}}\left( t \right)$ of the law \eqref{eq14} ensures the required control quality \eqref{eq13} for a system with the compensated uncertainty (i.e. when ${u_{ad}}\left( t \right) = {\theta ^{\rm{T}}}\Psi \left( t \right)$), the term ${u_{ad}}\left( t \right)$ is introduced to compensate for the uncertainty, and the nonlinear damping term ${u_{nd}}\left( t \right)$ is to guarantee the boundedness of all elements of ${e_{ref}}\left( t \right)$ for all bounded values of estimates $\hat \theta \left( t \right)$ of the uncertainty parameters.

Owing to \eqref{eq14}, the error equation obtained by subtraction of \eqref{eq13} from \eqref{eq12} is written as:
\begin{equation}\label{eq15}
{\dot e_{ref}}\left( t \right) = {A_{ref}}{e_{ref}}\left( t \right) + B\left( {{{\tilde \theta }^{\rm{T}}}\left( t \right)\Psi \left( t \right) + {u_{nd}}\left( t \right)} \right){\rm{,}}
\end{equation}
where $\tilde \theta \left( t \right) = \hat \theta \left( t \right) - \theta$.

In case of bounded parameter error $\tilde \theta \left( t \right) \in {\mathbb{L}_\infty }$, the nonlinear damping term $\kappa {B^{\rm{T}}}P{e_{ref}}\left( t \right){\Psi ^{\rm{T}}}\left( t \right)\Psi \left( t \right)$ ensures $\forall t \ge t_0^ +$  boundedness of the tracking error ${e_{ref}}\left( t \right) \in {\mathbb{L}_\infty }$ and, moreover, the steady-state value of such error ${e_{ref}}\left( {{t_{ss}}} \right)$ can be reduced by improvement of the value of $\kappa  > 0$ (see Lemma 2.26 \cite{b5}). Therefore, the main aim of adaptive control term $u_{ad}(t)$ is to ensure that, in the general case, $\tilde \theta \left( t \right) \in {\mathbb{L}_\infty }$ and, when $\Psi \left( t \right) \in {\rm{s\text{-}PE}}$, the following holds $\forall t \geq kT$:
\begin{equation}\label{eq16}
\mathop {{\rm{lim}}}\limits_{t \to \infty } \left\| {{e_{ref}}\left( t \right)} \right\| = 0{\rm{ }}\left( {{\rm{exp}}} \right){\rm{, }}\mathop {{\rm{lim}}}\limits_{t \to \infty } \left\| {\tilde \theta \left( t \right)} \right\| \le {\theta _{max }}{\rm{ }}\left( {{\rm{exp}}} \right).
\end{equation}

\section{Main Result}

To solve the problem \eqref{eq16} on the basis of the law \eqref{eq9}, first of all, a measurable static regression model of the form \eqref{eq8} with respect to a new vector of unknown parameters $\Theta \left( t \right)$ is obtained from the dynamic equation \eqref{eq15}. The way to derive this parametrization is described in the following statement.
\begin{proposition} On the basis of the measurable normalized signals:
    \begin{equation}\label{eq17}
\begin{array}{c}
z\left( t \right) =  - {\textstyle{{{B^{\rm{\dag }}}\left( {{e_{ref}}\left( t \right) - l{{\overline e}_{ref}}\left( t \right) - {A_{ref}}{{\overline e}_{ref}}\left( t \right)} \right) - \overline v\left( t \right)} \over {1 + {{\overline \Psi }^{\rm{T}}}\left( t \right)\overline \Psi \left( t \right)}}} = \\
 = {\theta ^{\rm{T}}}\overline \varphi \left( t \right) + {B^{\rm{\dag }}}{e^{ - l\left( {t - t_0^ + } \right)}}{\textstyle{{{e_{ref}}\left( {t_0^ + } \right)} \over {1 + {{\overline \Psi }^{\rm{T}}}\left( t \right)\overline \Psi \left( t \right)}}}{\rm{,}}\\
\overline \varphi \left( t \right) = {\textstyle{{\overline \Psi \left( t \right)} \over {1 + {{\overline \Psi }^{\rm{T}}}\left( t \right)\overline \Psi \left( t \right)}}}{\rm{, }}\\
\dot {\overline {\Psi}} \left( t \right) =  - l\overline \Psi \left( t \right) + \Psi \left( t \right){\rm{,\;}}\overline \Psi \left( {t_0^ + } \right) = {0_{m + n}}{\rm{,}}\\
{{\dot {\overline {e}}}_{ref}}\left( t \right) =  - l{{\overline e}_{ref}}\left( t \right) + {e_{ref}}\left( t \right){\rm{,\;}}{{\overline e}_{ref}}\left( {t_0^ + } \right) = {0_n}{\rm{,}}\\
\dot {\overline {v}}\left( t \right) =  - l\overline v\left( t \right) + {u_{ad}}\left( t \right) + {u_{nd}}\left( t \right){\rm{,\;}}\overline v\left( {t_0^ + } \right) = {0_m}
\end{array} 
\end{equation}
the procedures of dynamic extension
    \begin{equation}\label{eq18}
\begin{array}{c}
y\left( t \right){\rm{:}} = \mathfrak{H}\left\{ {\overline \varphi \left( t \right){z^{\rm{T}}}\left( t \right)} \right\}{\rm{, }}\\
\varphi \left( t \right){\rm{:}} = \mathfrak{H}\left\{ {\overline \varphi \left( t \right){{\overline \varphi }^{\rm{T}}}\left( t \right)} \right\}{\rm{, }}\\
d\left( t \right){\rm{:}} = \mathfrak{H}\left\{ {\overline \varphi \left( t \right){e^{ - l\left( {t - t_0^ + } \right)}}{\textstyle{{{{\left( {{B^{\rm{\dag }}}e_{ref}^{\rm{T}}\left( {t_0^ + } \right)} \right)}^{\rm{T}}}} \over {1 + {{\overline \Psi }^{\rm{T}}}\left( t \right)\overline \Psi \left( t \right)}}}} \right\},
\end{array}
\end{equation}
regularization
    \begin{equation}\label{eq19}
\begin{array}{c}
y\left( t \right) = \varphi \left( t \right)\theta  + d\left( t \right) = \\
\end{array}
\end{equation}
\vspace{-20pt}
\begin{gather*}
\begin{array}{c}
     = V\left( t \right)\Lambda \left( t \right){V^{\rm{T}}}\left( t \right)\theta  \pm V\left( t \right)\Xi \left( t \right){V^{\rm{T}}}\left( t \right)\theta  + d\left( t \right) = \\
 = V\left( t \right)\overline \Lambda \left( t \right){V^{\rm{T}}}\left( t \right)\theta  - V\left( t \right)\Xi \left( t \right){V^{\rm{T}}}\left( t \right)\theta  + d\left( t \right) = \\
 = \Phi \left( t \right)\theta  - V\left( t \right)\Xi \left( t \right){V^{\rm{T}}}\left( t \right)\theta  + d\left( t \right) = \\
 = \Phi \left( t \right)\Theta \left( t \right) + d\left( t \right),\\
\Xi \left( t \right) = \overline \Lambda \left( t \right) - \Lambda \left( t \right){\rm{,\;}}\Theta \left( t \right) = \theta  - {V_2}\left( t \right)V_2^{\rm{T}}\left( t \right)\theta {\rm{,}}\\
\overline \Lambda \left( t \right){\rm{:}} = \left\{ \begin{array}{l}
0,{\rm{\;if\;diag}}\left\{ {{{\overline \lambda }_1}\left( t \right){\rm{,}} \ldots {\rm{,}}{{\overline \lambda }_n}\left( t \right)} \right\} = \varepsilon {I_{n \times n}}\\
{\rm{diag}}\left\{ {{{\overline \lambda }_1}\left( t \right){\rm{,}} \ldots {\rm{,}}{{\overline \lambda }_n}\left( t \right)} \right\}{\rm{,\;otherwise}}
\end{array} \right.{\rm{,}}\\
{{\overline \lambda }_i}\left( t \right){\rm{:}} = \left\{ \begin{array}{l}
{\lambda _i}\left( t \right){\rm{,\;if\;}}{\lambda _i}\left( t \right) \ge \overline \varepsilon \\
\varepsilon {\rm{,\;if\;}}{\lambda _i}\left( t \right) < \overline \varepsilon 
\end{array} \right.{\rm{,\;}}i = \overline {1,n} {\rm{, }}
\end{array}
\end{gather*}
and mixing of the regressor:
    \begin{equation}\label{eq20}
\begin{array}{c}
\Upsilon \left( t \right){\rm{:}} = {{adj}}\left\{ {\Phi \left( t \right)} \right\}y\left( t \right){\rm{,\;}}\omega \left( t \right){\rm{:}} = {{det}}\left\{ {\Phi \left( t \right)} \right\}{\rm{, }}\\
\epsilon\left( t \right):= {{adj}}\left\{ {\Phi \left( t \right)} \right\}d\left( t \right){\rm{,}}
\end{array}
\end{equation}
the regression equation with measurable regressor  $\omega \left( t \right) \in \mathbb{R}$ and regressand $\Upsilon \left( t \right) \in {\mathbb{R}^{p \times m}}$ is obtained:
    \begin{equation}\label{eq21}
\Upsilon \left( t \right) = \omega \left( t \right)\Theta \left( t \right) + \epsilon\left( t \right){\rm{,}}
\end{equation}
where, when $\Psi \left( t \right) \in {\rm{s\text{-}PE}}$ with rank $r\left( t \right) \ge 1$, the inequality $\omega \left( t \right) \ge {\rm{min}}\left\{ {\lambda _{{\rm{min}}}^n\left( t \right){\rm{,\;}}{\varepsilon ^n}} \right\} \ge {\rm{min}}\left\{ {\mu^n{\rm{,\;}}{\varepsilon ^n}} \right\}  > 0$ holds $\forall t \ge kT$, and the new unknown parameters $\Theta \left( t \right)$ meet the indistinguishability condition:
    \begin{equation}\label{eq22}
{\Theta ^{\rm{T}}}\left( t \right)\Psi \left( t \right) - {\theta ^{\rm{T}}}\Psi \left( t \right) = {0_m}.
\end{equation}

Proof of Proposition 1 is postponed to Appendix.
\end{proposition}
As the condition \eqref{eq22} holds, the term ${\theta ^{\rm{T}}}\Psi \left( t \right)$ from \eqref{eq15} is substituted with ${\Theta ^{\rm{T}}}\left( t \right)\Psi \left( t \right)$ to obtain:
    \begin{equation}\label{eq23}
{\dot e_{ref}}\left( t \right) = {A_{ref}}{e_{ref}}\left( t \right) + B\left( {{{\tilde \Theta }^{\rm{T}}}\left( t \right)\Psi \left( t \right) + {u_{nd}}\left( t \right)} \right){\rm{,}}
\end{equation}
where $\tilde \Theta \left( t \right) = \hat \theta \left( t \right) - \Theta \left( t \right)$.

Owing to the boundedness of the new unknown parameters $\Theta \left( t \right)$, the goal \eqref{eq16} can be rewritten on the basis of \eqref{eq23} in an equivalent form:
    \begin{equation}\label{eq24}
\mathop {{\rm{lim}}}\limits_{t \to \infty } \left\| {{e_{ref}}\left( t \right)} \right\| = 0{\rm{ }}\left( {{\rm{exp}}} \right){\rm{, }}\mathop {{\rm{lim}}}\limits_{t \to \infty } \left\| {\tilde \Theta \left( t \right)} \right\| = 0{\rm{ }}\left( {{\rm{exp}}} \right).
\end{equation}

Let $\Psi \left( t \right) \in {\rm{s\text{-}PE}}$, then the aim is to analyze the properties of the system \eqref{eq23} from the point of view of conditions \eqref{eq16}, \eqref{eq24} under different assumptions concerning the rank $r\left( t \right)$ and basis ${V_2}\left( t \right)$ of the regressor $\varphi \left( t \right)$.
\begin{theorem}
    If  $\Psi \left( t \right) \in {\rm{s\text{-}PE}}$ and there exists the decomposition \eqref{eq3} with the time-invariant matrix  ${V_2}\left( t \right) \equiv {V_2}$  for the regressor  $\varphi \left( t \right)$ with time-invariant rank $r\left( t \right) \equiv r < n{\rm{,\;}}\linebreak \overline r\left( t \right) \equiv \overline r > 0$, then, when ${\gamma _0} > {\textstyle{1 \over {2\kappa }}}$, the adaptive law \eqref{eq9} ensures that \eqref{eq16} holds.
    
    Proof of Theorem 1 is given in Appendix.
\end{theorem}

Considering a second scenario, the rank $r\left( t \right)$ and nullspace ${V_2}\left( t \right)$ are often piecewise constant functions. In such case, the results of Theorem 1 describe local properties of the closed-loop system \eqref{eq15} over the time ranges when the rank and nullspace are time-invariant. The global properties of the errors $\tilde \theta \left( t \right){\rm{,\;}}{e_{ref}}\left( t \right)$  for the case under consideration are presented in the following theorem.

\begin{theorem}
    If $\Psi \left( t \right) \in {\rm{s\text{-}PE}}$, the rank $1 \le r\left( t \right) < n$ of the regressor $\varphi \left( t \right)$ is a piecewise-constant function:
\begin{gather*}
    \forall t \ge {t_{0}^+}{\rm{\;}}r\left( t \right) = \sum\limits_{{j_r} = 1}^{j_r^{\it max }} {{\Delta _{{j_r}}}h\left( {t - {t_{{j_r}}}} \right)},
\end{gather*}
and there exists the decomposition \eqref{eq3} with the piecewise-constant matrix ${V_2}\left( t \right)$:
\begin{gather*}
\forall t \ge {t_{0}^+}{\rm{\;}}{V_2}\left( t \right) = \sum\limits_{{j_V} = 1}^{j_V^{\it max }} {{\Delta _{{j_V}}}h\left( {t - {t_{{j_V}}}} \right)} {\rm{,}}
\end{gather*}
then, as the number of switches $j$ of the unknown parameters $\Theta \left( t \right) = \theta  + \sum\limits_{q = 1}^j {{\Delta _q}h\left( {t - {t_q}} \right)} $ is finite $j \le {j_{max}}$, the adaptive law \eqref{eq9} ensures that \eqref{eq16} holds when ${\gamma _0} > {\textstyle{1 \over {2\kappa }}}$.

Proof of Theorem 2 is presented in Appendix.
\end{theorem}

If the condition $\Psi \left( t \right) \in {\rm{s\text{-}PE}}$ does not hold, then, according to the procedures of the regularization \eqref{eq19} and mixing \eqref{eq20}, it follows that $\omega \left( t \right) = 0$ and, consequently, $ \dot {\tilde {\theta}} \left( t \right) = 0$. Therefore, owing to the nonlinear damping term $\kappa {B^{\rm{T}}}P{e_{ref}}\left( t \right){\Psi ^{\rm{T}}}\left( t \right)\Psi \left( t \right)$, we obtain (see Lemma 2.26 \cite{b5}) the desired boundedness 
\begin{center}
    $\tilde \theta \left( {t_0^ + } \right) \in {\mathbb{L}_\infty } \Rightarrow \tilde \theta \left( t \right) \in {\mathbb{L}_\infty } \Rightarrow {e_{ref}}\left( t \right) \in {\mathbb{L}_\infty }$. 
\end{center}

By similar reasoning, it is easy to check the boundedness of errors $\tilde \theta \left( t \right) \in {\mathbb{L}_\infty }{\rm{,\;}}{e_{ref}}\left( t \right) \in {\mathbb{L}_\infty }$ $\forall t \in \left[ {t_0^ + {\rm{;\;}}kT} \right)$ in the case when the condition $\Psi \left( t \right) \in {\rm{s\text{-}PE}}$ holds.

Thus, the adaptive control scheme \eqref{eq14}, \eqref{eq17}-\eqref{eq20}, \eqref{eq9} ensures the following properties:
\begin{enumerate}
    \item [--]${e_{ref}}\left( t \right) \in {\mathbb{L}_\infty }$ regardless of $\Psi \left( t \right)$;
    \item [--] $\Psi \left( t \right) \in {\rm{s\text{-}PE}} \Rightarrow \left\{ \begin{array}{l}
\mathop {{\rm{lim}}}\limits_{t \to \infty } \left\| {{e_{ref}}\left( t \right)} \right\| = 0{\rm{ }}\left( {{\rm{exp}}} \right){\rm{,}}\\
\mathop {{\rm{lim}}}\limits_{t \to \infty } \left\| {\tilde \Theta \left( t \right)} \right\| = 0{\rm{ }}\left( {{\rm{exp}}} \right){\rm{,}}\\
\mathop {{\rm{lim}}}\limits_{t \to \infty } \left\| {\tilde \theta \left( t \right)} \right\| \le {\theta _{max }}{\rm{ }}\left( {{\rm{exp}}} \right).
\end{array} \right.$
\item[--] the transient of  $\hat \theta \left( t \right)$ is of first-order type (each scalar parameter ${\hat \theta _i}\left( t \right)$ is adjusted using a separate first-order scalar differential equation), the rate of convergence of $\hat \theta \left( t \right)$ to $\Theta \left( t \right)$ can be adjusted with the help of  ${\gamma _0}$ value.
\end{enumerate}

Thus the main salient feature of proposed approach is that exponential convergence of tracking error is guaranteed under estimates convergence to new indistinguishable parameters.

\begin{remark}
In comparison with the adaptive laws from \cite{b44,b45}, the proposed approach guarantees the exponential convergence of the tracking error ${e_{ref}}\left( t \right)$ not only if the rank and nullspace of the regressor $\varphi \left( t \right)$ are constants (Theorem 1), but also if they are piecewise-constant functions (Theorem 2).
\end{remark}

\begin{remark}
    In contrast to the composite laws \cite{b22, b23, b24, b25, b26, b27, b28, b29, b30, b31}, the proposed adaptive control system \eqref{eq14}, \eqref{eq17}-\eqref{eq20}, \eqref{eq9} guarantees exponential stability without data stacks to store information about uncertainty, and, therefore, can be applied in case of time-varying parameters of uncertainty. In addition, the results of Theorem 2 allow one to assume that the developed system is capable of ensuring exponential stability under piecewise-constant uncertainty parameters with a finite number of switches. In this case, the jump change of the unknown parameters $\Theta(t)$ is caused not only by a change of the rank $r\left( t \right)$  and/or basis ${V_2}\left( t \right)$  of the regressor $\varphi \left( t \right)$, but also by a step-like change of the parameters $\theta $. The proof of exponential stability for the case of the piecewise-constant unknown parameters coincides with the proof of Theorem 2 exactly up to the fact that the exponentially decaying terms caused by violation of commutativity of filters \eqref{eq17}, \eqref{eq18} have been neglected.
\end{remark}
\begin{remark}
As $\omega(t) \geq min \{\mu^n,\;\varepsilon^n\} > 0$, the proposed system guarantees boundedness of ${e_{ref}}\left( t \right){\rm{,\;}}\tilde \Theta \left( t \right){\rm{,\;}}\tilde \theta \left( t \right)$ in the case of parameters $\theta$ variations or disturbances.
\end{remark}

\section{Numerical Experiments}
The following second-order plant has been chosen to conduct experiments:
  \begin{center}
$\begin{array}{c}
\left[ {\begin{array}{*{20}{c}}
{{{\dot x}_1}}\\
{{{\dot x}_2}}
\end{array}} \right] = \left[ {\begin{array}{*{20}{c}}
0&1\\
0&0
\end{array}} \right]\left[ {\begin{array}{*{20}{c}}
{{x_1}}\\
{{x_2}}
\end{array}} \right] + \left[ {\begin{array}{*{20}{c}}
0\\
1
\end{array}} \right]\left( {u - {\theta ^{\rm{T}}}\Psi \left( {x{\rm{,\;}}t} \right)} \right){\rm{,}}\\
{x_0} = {x_{0ref}} = {\left[ {\begin{array}{*{20}{c}}
{ - 1}&0
\end{array}} \right]^{\rm{T}}}{\rm{.}}
\end{array}$
\end{center}

The reference model \eqref{eq13} matrices, parameters of the filters \eqref{eq17}, \eqref{eq18}, regularization procedure \eqref{eq19}, control and adaptive laws \eqref{eq9}, \eqref{eq14} were chosen as:
$\begin{array}{c}
\begin{array}{*{20}{c}}
{k = l = 10,}\\
{\mu  = \varepsilon  = {\rm{1}}{{\rm{0}}^{ - 8}}},
\end{array}{\rm{\;}}\begin{array}{*{20}{c}}
{\kappa  = {\gamma _1} = 1,{\rm{\;}}{\gamma _0} = 10,}\\
{\bar \varepsilon  = {\rm{1}}{{\rm{0}}^{ - 17}}},
\end{array}{\rm{\;}}\begin{array}{*{20}{c}}
{{x_0} = {x_{0ref}}},\\
{Q = {I_2}},
\end{array}{\rm{}}\\
{K_x} = \left[ {\begin{array}{*{20}{c}}
{ - {\rm{5}}{\rm{.2915}}}&{ - {\rm{3}}{\rm{.2547}}}
\end{array}} \right]{\rm{,\;}}{K_r} =  - {\rm{5}}{\rm{.2915}}{\rm{.}}
\end{array}$

Three different uncertainties of the plant were used to conduct experiments.

\textbf{Case 1.} Functional uncertainty with time-invariant parameters and piecewise-constant rank of regressor.
  \begin{center}$
\begin{array}{c}
{\theta ^{\rm{T}}}{\rm{:}} = \left[ {\begin{array}{*{20}{c}}
{ - 1.75}&{0.5}
\end{array}} \right],\\
\Psi \left( {x{\rm{,\;}}t} \right){\rm{:}} = \left\{ \begin{array}{l}
\left[ {\begin{array}{*{20}{c}}
1&{ - 5\cos \left( {10t} \right)}
\end{array}} \right]{\rm{,\;if\;}}t \in \left[ {0{\rm{;\;5}}} \right],\\
\left[ {\begin{array}{*{20}{c}}
{\cos \left( {10t} \right)}&{ - 5\cos \left( {10t} \right)}
\end{array}} \right]{\rm{,\;}}\forall t \ge 5.
\end{array} \right.
\end{array}$
\end{center}

\textbf{Case 2.} Functional uncertainty with time-varying unknown parameters and constant rank of regressor.
  \begin{center}$
\begin{array}{c}
{\theta ^{\rm{T}}}{\rm{:}} = \left[ {\begin{array}{*{20}{c}}
{ - 1.75\sin \left( {25t} \right)}&{0.5}
\end{array}} \right],\\
\Psi \left( {x{\rm{,\;}}t} \right){\rm{:}} = \left[ {\begin{array}{*{20}{c}}
{\cos \left( {10t} \right)}&{ - 5\cos \left( {10t} \right)}
\end{array}} \right].
\end{array}$
\end{center}

\textbf{Case 3.} Wing-Rock-type nonlinearity with time-invariant parameters.
\begin{center}
$\begin{array}{c}
{\theta ^{\rm{T}}}{\rm{:}} = \left[ {\begin{array}{*{20}{c}}
{ - 22.22}&{23.74}&{ - 82.66}&{31.45}&{73.33}
\end{array}} \right],\\
\Psi \left( {x{\rm{,}}\;{\rm{ }}t} \right){\rm{:}} = {\left[ {\begin{array}{*{20}{c}}
{{x_1}}&{{x_2}}&{\left| {{x_1}} \right|{x_2}}&{\left| {{x_2}} \right|{x_2}}&{x_1^3}
\end{array}} \right]^{\rm{T}}}.
\end{array}$
\end{center}

Transients of states ${x_1}\left( t \right){\text{,\;}}{x_{1ref}}\left( t \right)$, uncertainty $\theta ^{{\text{T}}}\Psi \left( {x{\text{,}}\; {\text{}}t} \right)$, control signal $u_{ad}+u_{nd}$, parameters $\Theta \left( t \right)$ and obtained estimates $\hat \theta \left( t \right)$ for each of the above-mentioned cases are shown in Fig.1-3. Having analyzed the results, it was concluded that: 1) parameters estimates exponentially converged to indistinguishable parameters in all cases; 2) tracking error exponentially converged to zero; 3) proposed identifier was capable of tracking the unknown time-variant parameters. So, all theoretical results were validated.

\begin{figure}[thpb]\label{figure1}
\begin{center}
\includegraphics[scale=0.37]{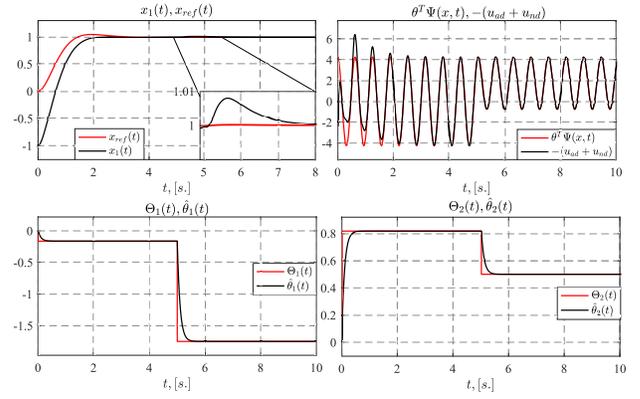}
\caption{{Transients for case 1.}} \end{center}
\end{figure}

\begin{figure}[thpb]\label{figure2}
\begin{center}
\includegraphics[scale=0.37]{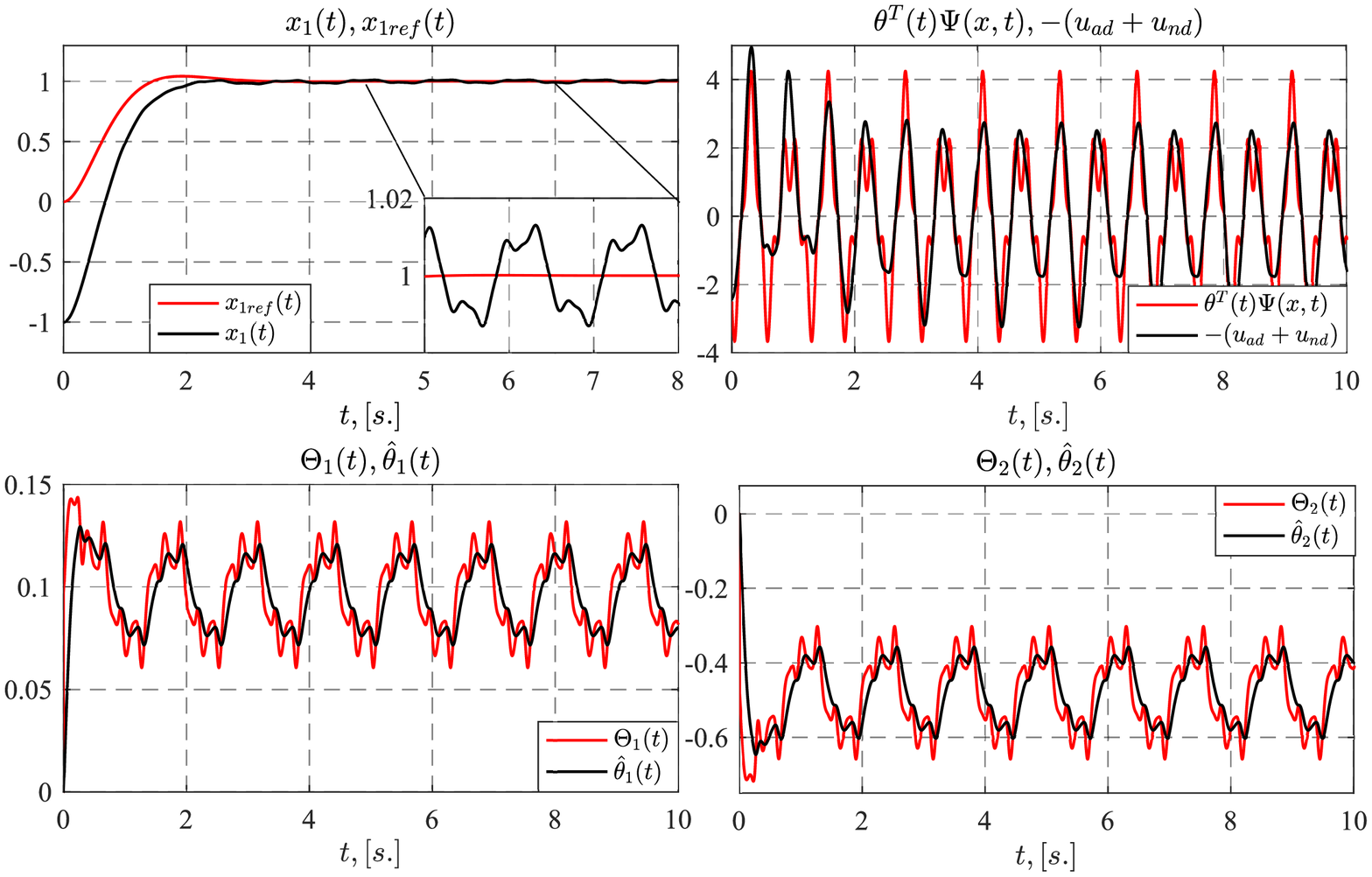}
\caption{{Transients for case 2.}} \end{center}
\end{figure}

\begin{figure}[thpb]\label{figure2}
\begin{center}
\includegraphics[scale=0.37]{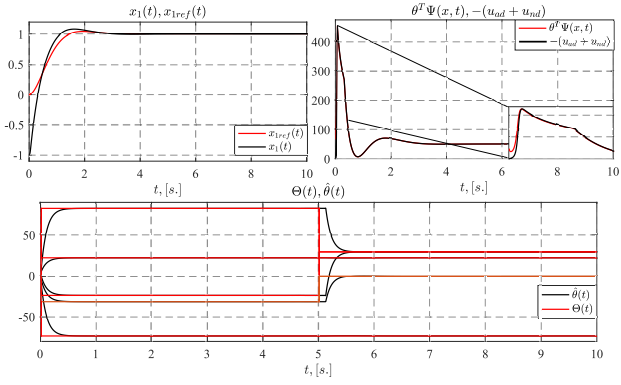}
\caption{{Transients for case 3.}} \end{center}
\end{figure}

\section{Conclusion and Future Work}
On the basis of the generalized procedure of dynamic regressor extension and mixing \cite{b47}, a new method of exponentially stable adaptive control was developed under the condition that the s-PE requirement was met. When such condition was not satisfied, the boundedness of the closed-loop system signals was ensured. The proposed adaptive control scheme was sensitive to changes of the unknown uncertainty parameters and robust to external disturbances. The properties of the system were proved analytically and validated in the course of the numerical experiments.

The scope of future research is to extend the obtained results to the class of systems with unknown matrix $B$ and to obtain the regression equation \eqref{eq21} with respect to indistinguishable parameters $\Theta(t)$ without using clipping regularization procedure \eqref{eq19} that is hard to be technically implemented.

\appendices

\renewcommand{\theequation}{A\arabic{equation}}
\setcounter{equation}{0}  

\section*{Appendix}
{\it{Proof of Proposition 1.}} The equation $\chi \left( t \right) = {e_{ref}}\left( t \right) -\linebreak- l{\overline e_{ref}}\left( t \right)$ is differentiated with respect to time:
    \begin{equation}\label{eqA1}
\begin{array}{c}
\dot \chi \left( t \right) = {l^2}{{\overline e}_{ref}}\left( t \right) - l{e_{ref}}\left( t \right) + \\
 + {A_{ref}}{e_{ref}}\left( t \right) + B\left( {{{\tilde \theta }^{\rm{T}}}\left( t \right)\Psi \left( t \right) + {u_{nd}}\left( t \right)} \right) = \\
 =  - l\chi \left( t \right) + {A_{ref}}{e_{ref}}\left( t \right) + \\
 + B\left( {{{\tilde \theta }^{\rm{T}}}\left( t \right)\Psi \left( t \right) + {u_{nd}}\left( t \right)} \right){\rm{,\;}}\chi \left( {t_0^ + } \right) = {e_{ref}}\left( {t_0^ + } \right).
\end{array}
\end{equation}

Having solved the equation \eqref{eqA1}, we have:
    \begin{equation}\label{eqA2}
\begin{array}{c}
\chi \left( t \right) = {e^{ - l\left( {t - t_0^ + } \right)}}{e_{ref}}\left( {t_0^ + } \right) + {A_{ref}}{{\overline e}_{ref}}\left( t \right) - \\
 - B{\theta ^{\rm{T}}}\overline \Psi \left( t \right) + B\overline v\left( t \right).
\end{array}
\end{equation}

The substitution of the function \eqref{eqA2} into the left-hand side of the equation \eqref{eq17} demonstrates that the right-hand side of \eqref{eq17} is valid.

Using the properties:
    \begin{equation}\label{eqA3}
\begin{array}{c}
{{adj}}\left\{ {\Phi \left( t \right)} \right\} = {{det}}\left\{ {\Phi \left( t \right)} \right\}{\Phi ^{ - 1}}\left( t \right){\rm{, }}\\
{\Phi ^{ - 1}}\left( t \right) = V\left( t \right){{\overline \Lambda }^{ - 1}}\left( t \right){V^{\rm{T}}}\left( t \right){\rm{,}}\\
{{adj}}\left\{ {\Phi \left( t \right)} \right\}\Phi \left( t \right) = {{det}}\left\{ {\Phi \left( t \right)} \right\}{I_n}{\rm{,}}
\end{array}
\end{equation}
and applying one after another the procedures of the dynamic extension \eqref{eq18}, regularization \eqref{eq19} and mixing \eqref{eq20}, the regression equation \eqref{eq21} with measurable regressand $\Upsilon \left( t \right)$ and scalar regressor $\omega \left( t \right)$, and unmeasurable exponentially vanishing disturbance $\epsilon\left( t \right)$ is obtained.

Following Lemma 6.8 \cite{b52}, when $\Psi \left( t \right) \in {\rm{s\text{-}PE}}$, it holds that $\overline \Psi \left( t \right) \in {\rm{s\text{-}PE}}$. In its turn, when $\overline \Psi \left( t \right) \in {\rm{s\text{-}PE}}$, we have $\overline \varphi \left( t \right) \in {\rm{s\text{-}PE}}$, so, according to Corollary 2, the inequality
\begin{center}
   $\omega \left( t \right)=\prod\limits_{i = 1}^n {{{\overline \lambda  }_i}\left( t \right)}  \ge {\rm{min}}\left\{ {\lambda _{{\rm{min}}}^n\left( t \right){\rm{,\;}}{\varepsilon ^n}} \right\} \geq \linebreak \geq {{min}}\left\{ {\mu^n{\rm{,\;}}{\varepsilon ^n}} \right\}> 0$  
\end{center}
also holds.

For the sake of convenience, the left-hand side of \eqref{eq22} is transposed, and the definition 
\begin{center}
   $ \Theta \left( t \right) = \theta  - V\left( t \right){\overline \Lambda ^{ - 1}}\left( t \right)\Xi \left( t \right){V^{\rm{T}}}\left( t \right)\theta$
\end{center}
is substituted into the obtained result:
    \begin{equation}\label{eqA4}
\begin{array}{c}
{\Psi ^{\rm{T}}}\left( t \right)\Theta \left( t \right) - {\Psi ^{\rm{T}}}\left( t \right)\theta \left( t \right) = \\
 - {\Psi ^{\rm{T}}}\left( t \right)V\left( t \right){{\overline \Lambda }^{ - 1}}\left( t \right)\Xi \left( t \right){V^{\rm{T}}}\left( t \right)\theta  = \\
 =  - {\Psi ^{\rm{T}}}\left( t \right){V_2}\left( t \right)V_2^{\rm{T}}\left( t \right)\theta .
\end{array}
\end{equation}

Because naturally, owing to the linear dependence of the matrix $\varphi \left( t \right)$ and vectors $\overline \varphi \left( t \right){\rm{,\;}}\overline \Psi \left( t \right)$ on $\Psi \left( t \right)$, the following implications hold:
    \begin{equation}\label{eqA5}
\begin{array}{l}
\varphi \left( t \right) \bot {V_2}\left( t \right)V_2^{\rm{T}}\left( t \right) \Leftrightarrow \overline \varphi \left( t \right) \bot {V_2}\left( t \right)V_2^{\rm{T}}\left( t \right) \Leftrightarrow \\
\overline \Psi \left( t \right) \bot {V_2}\left( t \right)V_2^{\rm{T}}\left( t \right) \Leftrightarrow \Psi \left( t \right) \bot {V_2}\left( t \right)V_2^{\rm{T}}\left( t \right)
\end{array}
\end{equation}
it follows from \eqref{eqA4} and \eqref{eqA5} that the indistinguishability condition \eqref{eq22} is met, which completes the proof.

{\it{Proof of Theorem 1.}} In the process of analysis of the errors ${e_{ref}}\left( t \right){\rm{,\;}}\tilde \theta \left( t \right)$ properties, without loss of rigor but to reduce complexity, we neglect the exponentially vanishing term $\epsilon\left( t \right)$.

First of all, the exponential convergence of the parameter error $\tilde \theta \left( t \right)$ to the goal set with bound ${\theta _{max }}$ is proved.

The quadratic form is introduced (time arguments are omitted for brevity):
     \begin{equation}\label{eqA6}
L = {\tilde \theta ^{\rm{T}}}\tilde \theta .
\end{equation}

The equation \eqref{eqA6} is differentiated along \eqref{eq9} to obtain:
     \begin{equation}\label{eqA7}
\begin{array}{c}
\dot L =  - 2{{\tilde \theta }^{\rm{T}}}\left( {\gamma \omega \left( {\omega \hat \theta  - \omega \theta  + \omega {V_2}V_2^{\rm{T}}\theta } \right)} \right) = \\
 =  - 2{{\tilde \theta }^{\rm{T}}}\gamma {\omega ^2}\tilde \theta  - 2{{\tilde \theta }^{\rm{T}}}\gamma {\omega ^2}{V_2}V_2^{\rm{T}}\theta .
\end{array}
\end{equation}

Having Proposition 1 and the definition of the parameter $\gamma$ from \eqref{eq9} at hand, the upper bound of the derivative \eqref{eqA7} $\forall t \ge kT$ is written as:
     \begin{equation}\label{eqA8}
\begin{array}{c}
\dot L \le  - 2{{\tilde \theta }^{\rm{T}}}{\textstyle{{{\gamma _0}} \over {{\omega ^2}}}}{\omega ^2}\tilde \theta  - 2{{\tilde \theta }^{\rm{T}}}{\textstyle{{{\gamma _0}} \over {{\omega ^2}}}}{\omega ^2}{V_2}V_2^{\rm{T}}\theta  \le \\
 \le  - 2{\gamma _0}{{\tilde \theta }^{\rm{T}}}\tilde \theta  - 2{\gamma _0}{{\tilde \theta }^{\rm{T}}}{V_2}V_2^{\rm{T}}\theta  \le \\
 \le  - 2{\gamma _0}{\left\| {\tilde \theta } \right\|^2} + 2{\gamma _0}\left\| {\tilde \theta } \right\|{\theta _{{{max}}}}.
\end{array}
\end{equation}

Here to obtain \eqref{eqA8} we take into account the fact that the spectral norm of the multiplier ${V_2}V_2^{\rm{T}}$ equals to one.

Assuming $a = \sqrt {2{\gamma _0}} \left\| {\tilde \theta } \right\|{\rm{,}}$  $b = \sqrt {2{\gamma _0}} {\theta _{{{max}}}}$ and using the inequality $- {a^2} + ab \le  - {\textstyle{1 \over 2}}{a^2} + {\textstyle{1 \over 2}}{b^2}$, it is obtained from \eqref{eqA8} that:
     \begin{equation}\label{eqA9}
\dot L \le  - {\gamma _0}L + {\gamma _0}\theta _{{{max}}}^2.
\end{equation}

Having solved $\forall t \ge kT$ the differential inequality \eqref{eqA9}, it is written:
     \begin{equation}\label{eqA10}
L \le {e^{ - {\gamma _0}\left( {t - kT} \right)}}{\left\| {\tilde \theta \left( {kT} \right)} \right\|^2} + \theta _{{{max}}}^2\int\limits_{kT}^t {{e^{ - {\gamma _0}\left( {t - \tau } \right)}}} d\tau.
\end{equation}

From \eqref{eqA10}, taking into consideration $L = {\left\| {\tilde \theta } \right\|^2}$ and the fact that $\forall c,{\rm{\;}}d$ the inequality $\sqrt {{c^2} + {d^2}}  \le \sqrt {{c^2}}  + \sqrt {{d^2}} $ holds, we have:
     \begin{equation}\label{eqA11}
\left\| {\tilde \theta \left( t \right)} \right\| \le {e^{ - 0,5{\gamma _0}\left( {t - kT} \right)}}\left\| {\tilde \theta \left( {kT} \right)} \right\| + {\theta _{max }},
\end{equation}
which proves that the inequality $\mathop {{\rm{lim}}}\limits_{t \to \infty } \left\| {\tilde \theta \left( t \right)} \right\| \le {\theta _{max }}$ from \eqref{eq16} holds.

When the premises of Theorem 1 are satisfied, the following is true: $\Theta \left( t \right) \equiv \Theta {\rm{,\;}}\dot \Theta  = 0,{\rm{\;}}\tilde \Theta \left( t \right) = \hat \theta \left( t \right) - \Theta {\rm{,}}$ then the function $L$ is chosen to prove the exponential convergence of ${e_{ref}}\left( t \right)$ to zero:
     \begin{equation}\label{eqA12}
\begin{array}{c}
L = e_{ref}^{\rm{T}}P{e_{ref}} + {\textstyle{1 \over 2}}{{\tilde \Theta }^{\rm{T}}}\tilde \Theta {\rm{,\;}}H = {\rm{blockdiag}}\left\{ {P{\rm{, }}{\textstyle{1 \over 2}}} \right\}{\rm{,}}\\
\underbrace {{\lambda _{{{min}}}}\left( H \right)}_{{\lambda _{\mathop{\rm m}\nolimits} }}{\left\| \xi  \right\|^2} \le V\left( {\left\| \xi  \right\|} \right) \le \underbrace {{\lambda _{{{max}}}}\left( H \right)}_{{\lambda _M}}{\left\| \xi  \right\|^2}.
\end{array}
\end{equation}

Owing to \eqref{eq9} and \eqref{eq23}, $\forall t \geq kT$ the derivative of the quadratic form \eqref{eqA12} is written as:
     \begin{equation}\label{eqA13}
\begin{array}{c}
\dot L = e_{ref}^{\rm{T}}\left( {A_{ref}^{\rm{T}}P + P{A_{ref}}} \right){e_{ref}} + \\
 + 2e_{ref}^{\rm{T}}PB{{\tilde \Theta }^{\rm{T}}}\Psi  - 2\kappa e_{ref}^{\rm{T}}PB{B^{\rm{T}}}P{e_{ref}}{\Psi ^{\rm{T}}}\Psi  - \\
 - {{\tilde \Theta }^{\rm{T}}}\dot {\tilde{ \Theta}}  =  - e_{ref}^{\rm{T}}Q{e_{ref}} + 2e_{ref}^{\rm{T}}PB{{\tilde \Theta }^{\rm{T}}}\Psi  - \\
 - 2\kappa {\left\| {e_{ref}^{\rm{T}}PB} \right\|^2}{\left\| \Psi  \right\|^2} - {{\tilde \Theta }^{\rm{T}}}{\gamma _0}\tilde \Theta .
\end{array}
\end{equation}

Applying the Young inequality:
     \begin{equation}\label{eqA14}
\left\| {e_{ref}^{\rm{T}}PB} \right\|\left\| {\tilde \Theta } \right\|\left\| \Psi  \right\| \le \kappa {\left\| {e_{ref}^{\rm{T}}PB} \right\|^2}{\left\| \Psi  \right\|^2} + {\textstyle{1 \over {4\kappa }}}{\left\| {\tilde \Theta } \right\|^2}
\end{equation}
the upper bound of the derivative \eqref{eqA13} is obtained:
     \begin{equation}\label{eqA15}
\begin{array}{c}
\dot L \le  - {\lambda _{min}}\left( Q \right){\left\| {{e_{ref}}} \right\|^2} + 2\kappa {\left\| {e_{ref}^{\rm{T}}PB} \right\|^2}{\left\| \Psi  \right\|^2} + \\+ {\textstyle{1 \over {2\kappa }}}{\left\| {\tilde \Theta } \right\|^2} - 
  2\kappa {\left\| {e_{ref}^{\rm{T}}PB} \right\|^2}{\left\| \Psi  \right\|^2} - {\gamma _0}{\left\| {\tilde \Theta } \right\|^2} \le \\
 \le  - {\lambda _{min}}\left( Q \right){\left\| {{e_{ref}}} \right\|^2} - \left( {{\gamma _0} - {\textstyle{1 \over {2\kappa }}}} \right){\left\| {\tilde \Theta } \right\|^2} \le \\
 \le  - \min \left\{ {{\textstyle{{{\lambda _{min}}\left( Q \right)} \over {{\lambda _{max }}\left( P \right)}}}{\rm{, 2}}{\gamma _0} - {\textstyle{1 \over \kappa }}} \right\}L.
\end{array}
\end{equation}

Therefore, when ${\gamma _0} > {\textstyle{1 \over {2\kappa }}}$, the tracking error ${e_{ref}}\left( t \right)$ exponentially converges to zero $\forall t \ge kT$, which completes the proof of Theorem.

{\it{Proof of Theorem 2.}}  Since function \eqref{eqA6} is differentiable even in the case of piecewise-constant unknown parameters $\Theta \left( t \right)$, then the proof of exponential convergence of the parameter error $\tilde \theta \left( t \right)$ to a bounded set with bound ${\theta _{max }}$  coincides with the one from Theorem 1.

The proof of Theorem second part is divided into two steps. In the first one, the exponential convergence $\tilde \Theta \left( t \right)$ is proved, and secondly, the exponential stability of ${e_{ref}}\left( t \right)$ is derived.

\textbf{Step 1.} The derivative of $\tilde \Theta \left( t \right)$ is written as:
     \begin{equation}\label{eqA16}
\dot {\tilde {\Theta}} \left( t \right) =  - \gamma \left( t \right){\omega ^2}\left( t \right)\tilde \Theta \left( t \right) - \sum\limits_{q = 1}^j {{\Delta _q}\delta \left( {t - {t_q}} \right){\rm{,}}} 
\end{equation}
where $\delta \left( {t - {t_q}} \right)$ is the Dirac function (unit impulse function at the time instant ${t_q}$).

Having solved $\forall t \ge kT$ \eqref{eqA16}, it is obtained:
     \begin{equation}\label{eqA17}
\begin{array}{l}
\tilde \Theta \left( t \right) = \phi \left( {t,kT} \right)\tilde \Theta \left( {kT} \right) - \\
{\rm{\hspace{65pt}}} - \int\limits_{kT}^t {\phi \left( {t,\tau } \right)\sum\limits_{q = 1}^i {{\Delta _q}\delta \left( {\tau  - {t_q}} \right)} d\tau } {\rm{,}}
\end{array}
\end{equation}
where $\phi \left( {t,\tau } \right) = {e^{ - \int\limits_\tau ^t {{\gamma _0}d\tau } }}$.

Using the sifting property of the Dirac function:
     \begin{equation}\label{eqA18}
\int\limits_{kT}^t {f\left( \tau  \right)\delta \left( {\tau  - kT} \right)d\tau }  = f\left( {kT} \right)h\left( {t - kT} \right){\rm{, }}\forall f\left( t \right){\rm{,}}
\end{equation}
we have from \eqref{eqA17}:
     \begin{center}\label{eqA19}
     $\begin{array}{c}
\left\| {\tilde \Theta \left( t \right)} \right\| \le \phi \left( {t,kT} \right)\left\| {\tilde \Theta \left( {kT} \right)} \right\| + \\
 + \sum\limits_{q = 1}^i {\phi \left( {t,{t_q}} \right)\left\| {{\Delta _q}} \right\|h\left( {t - {t_q}} \right)}  = \\
 = \underbrace {\left( {\left\| {\tilde \Theta \left( {kT} \right)} \right\| + \sum\limits_{q = 1}^i {\phi \left( {kT{\rm{, }}{t_q}} \right)\left\| {{\Delta _q}} \right\|h\left( {t - {t_q}} \right)} } \right)}_{\beta \left( t \right)}\phi \left( {t,kT} \right),
\end{array}$
\end{center}
where $\phi \left( {kT{\rm{,\;}}{t_q}} \right) = {\phi ^{ - 1}}\left( {{t_q},kT} \right) = {\phi ^{ - 1}}\left( {t,kT} \right)\phi \left( {t,{t_q}} \right) = \linebreak =\phi \left( {kT{\rm{,\;}}t} \right)\phi \left( {t,{t_q}} \right)$.

If the number of the parameters switches is finite $j \le \linebreak \le {j_{{{max}}}} < \infty $, then ${t_q}$ is also finite and the following upper bound holds:
     \begin{equation}\label{eqA20}
\begin{array}{l}
\beta \left( t \right) \le \left\| {\tilde \Theta \left( {kT} \right)} \right\| + \\
{\rm{           \hspace{35pt}             }} + \sum\limits_{q = 1}^{{j_{{{max}}}}} {\phi \left( {kT{\rm{, }}{t_q}} \right)\left\| {{\Delta _q}} \right\|h\left( {t - {t_q}} \right)}  = {\beta _{{\rm{max}}}}{\rm{,}}
\end{array}
\end{equation}

The parameter error $\tilde \Theta \left( t \right)$ exponential convergence follows immediately from the boundedness \eqref{eqA20} of $\beta \left( t \right)$:
     \begin{equation}\label{eqA21}
\left\| {\tilde \Theta \left( t \right)} \right\| \le {\beta _{max }}\phi \left( {t,kT} \right) = {\beta _{max }}{e^{ - {\gamma _0}\left( {t - kT} \right)}},
\end{equation}
as was to be proved in the first step, so the next aim is to analyze the tracking error ${e_{ref}}\left( t \right)$ behavior.

\textbf{Step 2.} Let the following quadratic form be introduced \linebreak $\forall t \geq kT$:
     \begin{equation}\label{eqA22}
     \begin{array}{c}
L = e_{ref}^{\rm{T}}P{e_{ref}}{\rm{ + }}{\textstyle{{\beta _{max }^2} \over 2}}{e^{ - 2{\gamma _0}\left( {t - kT} \right)}}{\rm{,}}\\H = {\rm{blockdiag}}\left\{ {P{\rm{,\;}}{\textstyle{{\beta _{max }^2} \over 2}}} \right\}{\rm{,}}\\
\underbrace {{\lambda _{{{min}}}}\left( H \right)}_{{\lambda _{\mathop{\rm m}\nolimits} }}{\left\| \xi  \right\|^2} \le V\left( {\left\| \xi  \right\|} \right) \le \underbrace {{\lambda _{{{max}}}}\left( H \right)}_{{\lambda _M}}{\left\| \xi  \right\|^2}{\rm{,}}
\end{array}
\end{equation}
where $\xi \left( t \right) = {\left[ {\begin{array}{*{20}{c}}
{e_{ref}^{\rm{T}}\left( t \right)}&{{e^{ - {\gamma _0}\left( {t - kT} \right)}}}
\end{array}} \right]^{\rm{T}}}.$

The derivative of \eqref{eqA22} is written as:
     \begin{equation}\label{eqA23}
\begin{array}{c}
\dot L = e_{ref}^{\rm{T}}\left( {A_{ref}^{\rm{T}}P + P{A_{ref}}} \right){e_{ref}} + \\ + 2e_{ref}^{\rm{T}}PB{{\tilde \Theta }^{\rm{T}}}\Psi
 - 2\kappa e_{ref}^{\rm{T}}PB{B^{\rm{T}}}P{e_{ref}}{\Psi ^{\rm{T}}}\Psi  - \\ - {\gamma _0}\beta _{max }^2{e^{ - 2{\gamma _0}\left( {t - kT} \right)}} = \\ = - e_{ref}^{\rm{T}}Q{e_{ref}} + 2e_{ref}^{\rm{T}}PB{{\tilde \Theta }^{\rm{T}}}\Psi  - \\
 - 2\kappa {\left\| {e_{ref}^{\rm{T}}PB} \right\|^2}{\left\| \Psi  \right\|^2} - {\gamma _0}\beta _{max }^2{e^{ - 2{\gamma _0}\left( {t - kT} \right)}}.
\end{array}
\end{equation}

Having the inequality \eqref{eqA14} at hand, the upper bound of the derivative \eqref{eqA23} is obtained:
     \begin{equation}\label{eqA24}
\begin{array}{c}
\dot L \le  - {\lambda _{min}}\left( Q \right){\left\| {{e_{ref}}} \right\|^2} + 2\kappa {\left\| {e_{ref}^{\rm{T}}PB} \right\|^2}{\left\| \Psi  \right\|^2} + 
\end{array}
\end{equation}
$\begin{array}{c}
+{\textstyle{1 \over {2\kappa }}}{\left\| {\tilde \Theta } \right\|^2} 
 - 2\kappa {\left\| {e_{ref}^{\rm{T}}PB} \right\|^2}{\left\| \Psi  \right\|^2} - {\gamma _0}\beta _{max }^2{e^{ - 2{\gamma _0}\left( {t - kT} \right)}} \le \\
 \le  - {\lambda _{min}}\left( Q \right){\left\| {{e_{ref}}} \right\|^2} + {\textstyle{1 \over {2\kappa }}}{\left\| {\tilde \Theta } \right\|^2} - {\gamma _0}\beta _{max }^2{e^{ - 2{\gamma _0}\left( {t - kT} \right)}}.
\end{array}$

The substitution of \eqref{eqA21} into \eqref{eqA24} yields:
     \begin{gather*}
\begin{array}{c}
\dot L \le  - {\lambda _{min}}\left( Q \right){\left\| {{e_{ref}}} \right\|^2} + {\textstyle{1 \over {2\kappa }}}\beta _{max }^2{e^{ - 2{\gamma _0}\left( {t - kT} \right)}} - \\
 - {\gamma _0}\beta _{max }^2{e^{ - 2{\gamma _0}\left( {t - kT} \right)}} \le  - \min \left\{ {{\textstyle{{{\lambda _{min}}\left( Q \right)} \over {{\lambda _{max }}\left( P \right)}}}{\rm{, 2}}{\gamma _0} - {\textstyle{1 \over \kappa }}} \right\}L.
\end{array}
     \end{gather*}

Therefore, when ${\gamma _0} > {\textstyle{1 \over {2\kappa }}}$, the tracking error ${e_{ref}}\left( t \right)$ exponentially converges to zero $\forall t \ge kT$, which completes the proof of Theorem.

\end{document}